# Evaluation of investigational paradigms for the discovery of non-canonical astrophysical phenomena

Caitlyn A. K. Singam[1], Jacob Haqq-Misra[2], Amedeo Balbi[3], Alexander M. Sessa[4], Beatriz Villarroel[5,6], Gabriel G. De la Torre[7], David Grinspoon[8], Hikaru Furukawa[9], Virisha Timmaraju[10]

***Abstract:*** *Non-canonical phenomena – defined here as observables which are either insufficiently characterized by existing theory, or otherwise represent inconsistencies with prior observations – are of burgeoning interest in the field of astrophysics, particularly due to their relevance as potential signs of past and/or extant life in the universe (e.g. off-nominal spectroscopic data from exoplanets). However, an inherent challenge in investigating such phenomena is that, by definition, they do not conform to existing predictions, thereby making it difficult to constrain search parameters and develop an associated falsifiable hypothesis.*

*In this Expert Recommendation, the authors evaluate the suitability of two different approaches – conventional parameterized investigation (wherein experimental design is tailored to optimally test a focused, explicitly parameterized hypothesis of interest) and the alternative approach of anomaly searches (wherein broad-spectrum observational data is collected with the aim of searching for potential anomalies across a wide array of metrics) – in terms of their efficacy in achieving scientific objectives in this context. The authors provide guidelines on the appropriate use-cases for each paradigm, and contextualize the discussion through its applications to the interdisciplinary field of technosignatures (a discipline at the intersection of astrophysics and astrobiology), which essentially specializes in searching for non-canonical astrophysical phenomena.*

## 1. Introduction

Non-canonical phenomena (abbreviated here as NCPs), particularly with the rise of quantum physics, have acquired newfound significance in physics and astrophysics due to their potential to provide insight into the validity, or lack thereof, of existing theories (for instance, the perihelion precession of Mercury, which was considered anomalous prior to its explanation via Einstein's theory of general relativity in 1916[1,2]), and to provide groundwork for formulating novel explanations[3,4]. The Zeeman effect[5] (prior to the characterization of electron spin in the 1900s) and the anomalous Hall effect[6] – both fundamental phenomena in physics – are two such examples of NCPs of import to the physical sciences. Additionally, NCPs have taken on extra significance in an astrophysics/astrobiology context as potential indicators of either past or extant biotic entities (e.g. the anomalous finding of high levels of phosphine on Venus, reported in *Nature Astronomy* in September 2020[7]) given that astrobiological entities would – due to a dearth of characterizing data and consensus theory – likely produce signals that would be categorized as NCPs.

---

[1] Bioengineering Department & Institute for Systems Research, University of Maryland, College Park, MD, USA. Corresponding author (csingam@umd.edu).
[2] Blue Marble Space Institute of Science, Seattle, WA, USA
[3] Dipartimento di Fisica, Università di Roma Tor Vergata, Roma, Italy
[4] Georgia Institute of Technology, Atlanta, GA, USA
[5] KTH Royal Institute of Technology and Stockholm University, Stockholm, Sweden
[6] Instituto de Astrofísica de Canarias, Tenerife, Spain
[7] University of Cadiz, Cadiz, Spain
[8] Planetary Science Institute, Tucson, AZ, USA
[9] Arizona State University, Tempe, Arizona, USA
[10] Jet Propulsion Laboratory, California Institute of Technology, Pasadena, CA, USA

Once recognized, individual NCPs can be studied utilizing standard investigational methodologies much in the same way as canonical phenomena; however, the varied nature of NCPs (due to the fact that the definition of an 'anomaly' or 'non-canonical' varies based on context) has generally thwarted attempts to implement systematic searches for non-canonicals of interest. Most NCPs are therefore identified serendipitously, as unexpected anomalies in hypothesis-driven experiments concerning canonical phenomena, rather than from searches for the NCPs themselves. As NCPs become of increasing scientific interest, though – particularly as 'points of significance' in model selection and development[8] – this passive approach to NCP discovery may no longer meet scientific needs. For NCP-dependent disciplines such as the burgeoning field of technosignatures, which use astrophysics methods and data to conduct to achieve their objectives, an inability to perform structured astrophysics NCP searches means has already proved limiting in the scope of investigations that can be performed (Box 1).

There is thus a cogent need for a clear investigational paradigm for structured NCP discovery. This Expert Recommendation seeks to address this need by comparing the relative suitability of two different investigational paradigms – parameterized investigation, and broad-spectrum anomaly searches – for structured NCP discovery and characterization, and by presenting a generic (use-case agnostic) NCP discovery investigational lifecycle for astrophysics research. This paper also presents an evaluation of how technosignatures research can implement the proposed paradigm, as an example of its utility for highly NCP-dependent research efforts.

---

**Box 1 – NCP-dependency of biosignature/technosignature research**

The field of technosignatures is a newly emerging subdiscipline of astrobiology. Like biosignatures, which are substances or observational phenomena that provide scientific evidence of past/extant life, technosignatures are scientific evidence of artificial constructs generated by any type of biotic entities. Atmospheric emissions (e.g. nitrogen dioxide and chloroflurocarbons on Earth, which are detectable by remote instruments[9]) and artificial/non-geothermal heat signatures are generally considered to be examples of such technosignatures.

A major challenge of the field is that existing data on biosignatures and technosignatures is currently limited to the environment of Earth, for a sample size of 1. There is therefore an inherent statistical risk in inferring the characteristics of potential biosignatures/technosignatures using existing canonical datasets and theoretical models; the fact that the signatures of biological life are highly-context dependent because of the adaptive nature of biotic entities only compounds this risk. Given that current investigational paradigms call for parametrization of the search target(s) prior to experiment/instrument development, most investigations into biosignatures and technosignatures have thus been limited to searching for canonical phenomena consistent with observations on Earth or projections of future Earth scenarios.

However, the inconsistency between most planetary environments and Earth's own biomes means that even canonical entities, if they existed on other planets, would likely produce non-canonical signatures due to differences in how biological systems would be forced to interact with their environments[10]. Thus, the search for biosignatures and technosignatures is fundamentally tied to the identification of new NCPs, which are the primary avenues by which to expand search parameters.

## 2. Definition of NCPs

For the purposes of this paper, NCPs are defined as observed entities or events which are inconsistent with predictions made by the scientific canon (i.e. existing empirically-supported theories and prior verified observations). The level of inconsistency between canon and an NCP can be quantified based on a comparison of quantitative observed versus expected values for metrics of interest, relative to the amount of random variation seen in the data-set. Based on the Bienaymé–Chebyshev inequality[11][12], which describes the probability of an observed population value $X$ falling beyond $k$ standard deviations of an expected value µ by random chance [13],

$$Pr\ (|X - \mu| \geq k\ ) \ \leq \ \frac{\sigma^2}{k^2}$$

3 sigma is the generally accepted threshold of significance for a phenomenon to potentially be an NCP[14]. The more stringent 5 sigma significance threshold (which represents statistical confidence that only 4% of observations, at most, will fall beyond $k = 5$ standard deviations of the expected value µ) that is considered the gold standard for identifying discoveries in particle physics[15] is applied here for conclusively identifying an observation as a genuine NCP.

## 3. Investigational paradigms
### 3.1. Parameter-based studies

The conventional approach to conducting experimental studies, particularly in the physical sciences, is to employ a highly parameterized investigation wherein one formulates a testable hypothesis relative to a handful of metrics of interest. This approach is extremely effective in evaluating a known phenomenon in increased depth, especially when one tailors the experimental setup to suit the hypothesis being tested. However, parameter-based studies are not well-suited to identifying phenomena which are not understood well enough to be characterized; developing a specific hypothesis requires a degree of understanding of the phenomena of interest which is difficult to attain in the case of NCPs. For instance, in the case of astrobiological investigations, one cannot search for the full gamut of possible non-canonical life forms without the ability to constrain the characteristics of life on other planets (Box 1).

Another limitation of parameter-based studies is the risk of NCPs being disregarded as outliers; many studies employ techniques such as the Q test[16] or Grubbs' test[17] for outlier exclusion. Whilst such practices are not as common in the physical sciences due to concerns about compromising data integrity, etc., the practice is nonetheless reflective of how parameter-based studies tend to be optimized for analyzing phenomena that are pertinent to supporting or rejecting the hypothesis of interest. The strength of parameterized investigations lies in characterizing phenomena once they have been identified; performing the large-scale data analysis needed to initially recognize a previously unknown NCP is best done by other means.

### 3.2. Anomaly-based studies

Anomaly-based studies are a subset of exploratory research paradigms, guided by the objective of formulating meaningful hypotheses without any prior assumptions about the collected data. Rather than seeking to confirm or reject a hypothesis regarding the existence of certain phenomena, anomaly-based studies aim to find phenomena which do not conform to the behaviors characterized by existing canon.

The cause for data anomalies is typically attributable to either some form of error (experimental, instrument, human, etc.) or to novel phenomena (NCPs). Bayesian statistical methods are typically an effective means of separating NCP-associated data points from the rest of a particular dataset, though performance is significantly improved if one works with a larger sample size.

In astrophysics, the "cosmic haystack" of data collected by several modern digital sky surveys provides a vast parameter space, defined by several observable dimensions, and is a good candidate for use in a systematic search for NCPs. However, anomaly-based studies pose unique challenges; the lack of a formal, fully constrained hypothesis means that subtle anomalies (i.e. minor inconsistencies between observed data and prior predictions) may be overlooked. Thus, exploratory research and anomaly detection paradigms are best emphasized for NCP identification, with characterization efforts relegated to parameter-based studies.

### 3.3. Dogmatic context

The two main approaches to NCP identification – parameterized investigations centered around narrowly-scoped hypotheses, and broader-scope anomaly searches – are reflective of two different scientific dogmas. Karl Popper's model of science[18][19] – consistent with parameterized investigations – described the state of science at any given time as consisting of a large set of hypotheses that have been rejected by experiments and a smaller set of tentative hypotheses that represent the best set of ideas that have not yet been falsified. Robust evidence, confirmed by multiple experiments, is sufficient to reject a previously accepted hypothesis in favor of a new one that shows better agreement with the available data. Experiments that provide confirmation of existing hypotheses are also useful, but they do not "verify" the hypothesis as being true in the sense of logical positivism. Instead, positive tests of hypothesis provide supporting evidence of the robustness of the hypothesis, or set of hypotheses, being tested. In many cases Bayesian methods can be used to quantify the extent to which hypotheses are congruent with data [20], although such methods still cannot establish the epistemological truth of a hypothesis.

In contrast, Thomas Kuhn[21] – whose ideas support the anomaly search-based approach to NCP identification – developed a description of progress in science as alternating between periods of "normal" inquiry and the onset of "crisis." Kuhn's model recognized that scientific revolutions have historically featured substantial challenge to the status quo, such as the assertion of heliocentrism against geocentric models or the recognition that Newtonian mechanics becomes inadequate at relativistic or quantum scales. In a period of normal science, Kuhn describes scientists as working within a known paradigm to solve puzzles that arise within the paradigm. However, the recognition of problems that are inconsistent or challenging to the paradigm bring science (or a particular discipline) to a state of crisis, in which the paradigm has been shown to be problematic but no alternative is available. Eventually, a new paradigm emerges that remains consistent with all available data, and science returns to a state of normal inquiry. Under Kuhn's model, the process of hypothesis testing remains an adequate description of normal science, but the recognition of a crisis within a scientific discipline requires exploratory methods that seek to detect anomalies or other inconsistencies with the prevailing paradigm. This approach does not necessarily require that an alternative hypothesis or paradigm be immediately provided, as the phase of crisis is marked by the accumulation of anomalous data that draws attention to underlying problems with the dominant paradigm.

It is worth noting that Popper's and Kuhn's approaches are not mutually exclusive from one another and are in fact, to a great extent, complementary, with anomaly-driven investigation being suitable to times of 'crisis' when Popper's parameterized hypothesis-driven approach may not suffice. In the case of NCP identification, which represents a balance of what Kuhn might consider 'normal' and 'crisis' inquiry - it is thus justified to develop an approach that deploys both anomaly searches and parameterized investigations in the contexts that suit each method best.

## 4. Proposed paradigm
### 4.1. Generic case

The paradigms presented above are best employed in conjunction with one another, as part of a multi-stage program. Exploratory investigations, which are well-suited to broad searches for anomalous phenomena, are best employed in the early phases of NCP research (i.e. it is difficult to place constraints on NCP characteristics due to limited prior knowledge). Findings from such investigations can subsequently be examined via parameterized, narrowed-scope experiments, where it is possible to formally construe a testable hypothesis based on observations of the newly identified NCP and place constraints on the phenomena being studied based on existing observations/data. Thus, it is possible to cater instrument design, investigational targets, etc. for the parameter-based follow-up studies to better support characterization of the NCP of interest based on hypotheses about its nature.

The utility of this multi-stage investigational life cycle is best demonstrated via a generic hypothetical example. Consider, for instance, the plot shown in Figure 1a of the following equation:

$$M = p_1 + p_2 + p_3 + N$$

where $p_1$, $p_2$, and $p_3$ are environmental phenomena which affect measured values of an observable metric $M$ at any point in time $t$, and $N$ is a variable representing noise. (One can contextualize $M$ as being the environmental levels of a compound of interest in an exoplanet's atmosphere, etc.).

In the case of Figure 1, the plot was generated using the following equations for $p_1$, $p_2$, and $p_3$:

$$p_1 = \sin(t) + 1$$

$$p_2 = \cos\left(\left(\frac{t}{4}\right)^2\right) + 1$$

$$p_3 = \sin\left(\frac{t}{2}\right) + \frac{\sin\left(\frac{t}{4}\right)}{50 * \sin(t)}$$

$N$ was generated by drawing random samples from a Gaussian distribution centered at 0 with a standard deviation of 0.25. Figures 1b-1e shows the time series plots for each of the terms composing $M$.

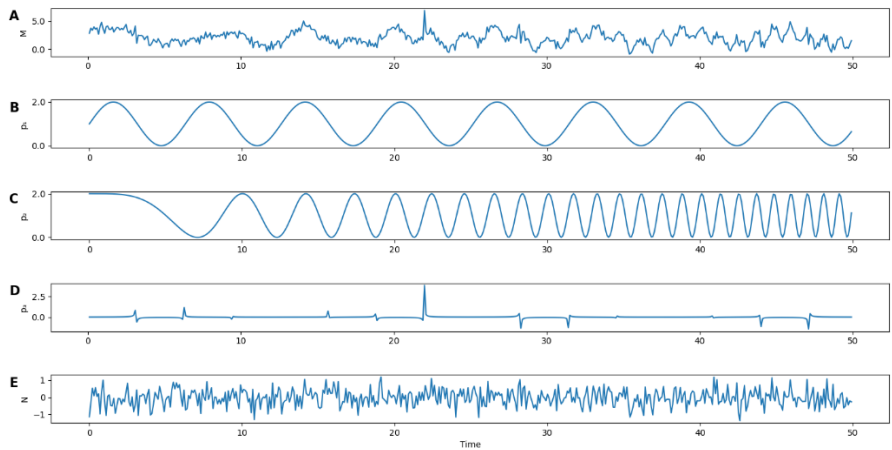

**Figure 1. Generic hypothetical dataset for analysis by parameterized and anomaly-driven approaches. (a) The metric (*M*) is a time-series generated by adding the effects of *p₁*, *p₂*, and *p₃***

together with Gaussian noise. (b) Canonical, known phenomenon 1 ($p_1$), plotted here as a separate time-series, is one of the terms used to generate *M*. (c) Canonical phenomenon 2 ($p_2$), plotted here as a separate time-series, is another one of the terms used to generate *M*, though the investigator in this scenario is uncertain of this fact. (d) Noncanonical phenomenon 3 ($p_3$), plotted here as a separate time-series, is also used to generate *M*, though this fact is unknown to the investigators in this scenario. (e) The Gaussian noise added to $p_1$, $p_2$, and $p_3$ to generate *M*.

Let it be assumed further in this scenario that $p_1$ is a canonical phenomenon (e.g. a geophysical process on the planet) that is known to be contributing to the value of *M*, that $p_2$ is a canonical phenomenon is known to potentially affect *M*, but has not yet been shown to occur in the environment of interest (e.g. a biotic phenomenon which affects the value of *M* on Earth, but has not yet been shown to exist on the planet being studied) and, and that $p_3$ is a wholly unknown canonical phenomenon (e.g. an NCP associated with a biotic entity dissimilar to those observed on Earth).

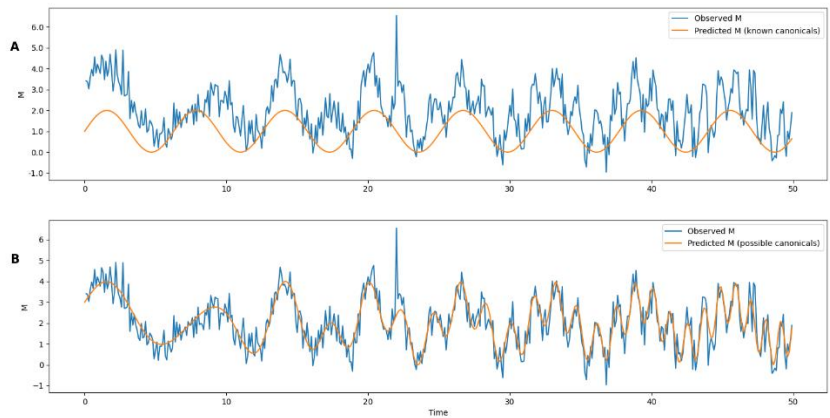

**Figure 2. Parameterized, constrained hypothesis-driven approach to analyzing metric M. (a) The prediction for the value of M based solely on known canonical $p_1$ is insufficient to explain the variance seen in M. (b) A prediction based on canonical $p_1$ and possibly relevant canonical $p_2$ shows significantly higher efficacy in explaining the variance observed in M.**

If one takes a traditional, parameterized constrained-hypothesis approach, one would approach the data with the hypothesis that the observed value of M is a combination of $p_1$ and $p_2$. Visualizing the data (Figure 2a) yields clear evidence that $p_1$ insufficiently characterizes the behavior of *M*; however, the addition of $p_2$ (Figure 2b) drastically increases the fidelity of the model relative to the observed value of *M*. Performance of an F-test yields statistical evidence that the combined $p_1 – p_2$ model is effective in explaining the variance observed in M, and the study concludes that M is influenced by both $p_1$ and $p_2$.

Note that this approach requires the investigator to have prior knowledge of the behavior of the secondary phenomenon, $p_2$, which is being tested for. This is knowledge that investigators do not necessarily have when searching for novel NCPs, making the conventional approach ineffective and somewhat limiting in its range of potential targets. In contrast, the proposed investigational lifecycle paradigm is a two-phased approach. Unlike the traditional approach, which seeks to determine whether or not a specific, well-characterized phenomenon is the cause of an observed effect, the proposed paradigm starts with a hypothesis-agnostic approach, parsing multiple metrics for anomalies that deviate from the remainder of the dataset (or from other similar datasets). Apart from the assumption that NCPs will cause anomalous perturbations in observable metrics, there is no assumption made on the characteristics of the NCP (e.g. when it will occur in the data, etc.)

In the case of the generic case presented here, metric *M* has an anomalous reading at approximately t=22. This anomaly is detectable based on its deviation from the mean; the presence of the anomaly thus prompts observation of metric *M* over an extended time frame for evidence of possible recurrence of the anomaly, as seen in Figure 3a. (Non-recurring, non-reproducible anomalies are more likely to be indicative of the observed anomaly being a product of instrument artifact or experimental error rather than a novel phenomenon, but are nonetheless still potentially warrant further investigation).

Once the anomaly is identified, one can attempt to correct for the effect of any known phenomena (in this case, $p_1$), and perform conventional hypothesis to determine if any other canonical phenomena (in this case, $p_2$) affect *M*. The effect of these canonicals can then be controlled for when analyzing the M dataset – done here by subtracting the $p1$ and $p_2$ time series from M - yielding a partial characterization of the potential NCP (Figure 3b). Using this partial characterization, investigators can then perform additional studies to further parametrize the new NCP ($p_3$) and verify results by comparing revised models (that include the NCP) against observations using statistical methods (Figure 3c).

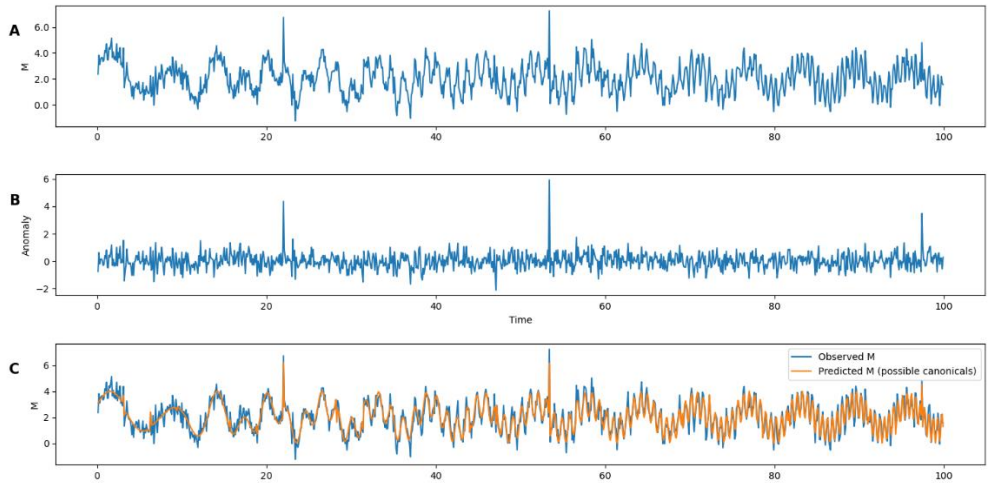

**Figure 3. Analysis of metric *M* via the proposed paradigm. (a) Extended-duration analysis of metric *M*. (b) Partial anomaly characterization by controlling for the effect of known canonicals on *M*. (c) Revised model for *M*, inclusive of NCP $p_3$.**

### 4.2. Recommendation

This proposed investigational lifecycle can be generalized into the diagram seen in Figure 4. As shown in the diagram, it is to be expected that some of the canonical phenomena observed during an anomaly search will be nominated for additional study (via parameterized, constrained investigations) due to their high science value, even if they are not nominally NCPs. These additional studies may lead to insight about potential NCPs peripheral to existing canonical phenomena which did not meet the threshold of detection in prior studies, or may otherwise be of interest due to other scientific priorities. The datasets deemed to be of medium and low priority also potentially warrant study, but are not considered to be of immediate priority for investigation and may be studied at a subsequent point in time. Nonetheless, it is recommended that the medium and low priority data that is collected during investigations be archived so as to aid future NCP research efforts.

Compared to conventional methodologies (which follow the nominal-high science value-secondary studies path in the Sankey diagram shown in Figure 4, and which only study a small fraction of the potential datasets of interest), the proposed investigational lifecycle is significantly more effective in identifying which metrics are likely to yield higher scientific value upon subsequent investigation. In effect, while both the conventional and proposed methodologies are reliant on NCP-associated statistical priors (i.e. data that indicates an NCP is likely to be present), the conventional methodology is reliant on theorized priors – generated based on canonical theory, investigator assumptions, and prior data from other contexts – the proposed paradigm uses the anomaly-search phase as an opportunity to develop a tailored statistical prior (i.e. the empirical data collected on the anomaly). Thus, when an anomaly is identified by the proposed paradigm, one is able to tailor parameterized investigations to a phenomenon which has been empirically observed in the context of interest, rather than to a theoretical prediction, thereby increasing the likelihood of subsequent characterization studies being able to yield more specific data.

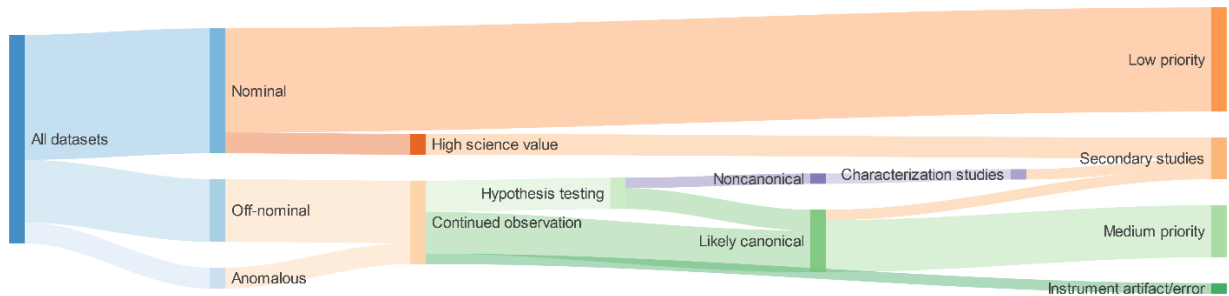

**Figure 4. A diagram depicting the NCP identification investigational life cycle. Separation of datasets into nominal/off-nominal/anomalous categories is done via anomaly searches, whereas characterization studies and secondary follow-up studies are intended to be parameterized, constrained investigations.**

## 5. Case study: technosignatures
### 5.1. Case study context

The field of technosignatures is an emerging NCP-dependent discipline dedicated to searching for artificial constructs produced by life. As described in Box 1, a major challenge facing the technosignature community has been identifying an appropriate investigational paradigm for systematic technosignature identification, especially given the lack of non-Earth-based technosignature examples.

The lack of existing data on definitive exoplanetary technosignatures has meant that there has been a dire need for the definition of criteria, however broad, to define the scope of any specific technosignatures observational campaign, or analysis of existing data. Such criteria would also aid in evaluating the outcome of technosignature-relevant observations, both in the case of negative results and in the case of candidate signatures or anomalies, especially when the data being analyzed is from pre-existing datasets (which were not necessarily explicitly designed for technosignatures research).

The following sections serve as a case study in the scientific challenges of studying NCPs, and showcase the suitability of the proposed investigational lifecycle in guiding the formation of an operational definition for technosignatures and potentially directing the future of NCP-identification research.

### 5.2. Challenges of technosignature research

As previously discussed, one of the largest challenges in technosignature research is the lack of a robust, parameterized definition of technosignatures. While it technosignatures can be defined as 'artificial constructs produced by life', it is unclear how, precisely, one can use that definition to define search parameters. For example, if one were to observe unusual quantities of abiotic compounds in the atmosphere of a planet, it is difficult to ascertain if those compounds are artificially produced by life or if there are unknown natural processes on that planet to produce such compounds. The potential detection of abnormal levels of phosphine on Venus is a typical example.) The same challenges exist for biosignature research, the parent discipline of technosignatures – and another NCP-driven field – since existing data on life is limited to that on Earth (for a sample size of n=1).

The current, canonically-based approach is to constrain the definition of technosignature in a narrow sense, and parametrize it based on existing Earth-based observational data (from Earth-orbiting satellites, etc.) since the relationship between observable metrics and causative agents is well-understood in that context. However, given that exoplanetary technosignatures are likely to be NCPs which do not align with canonical Earth-centric technosignatures, an alternate investigational paradigm is needed.

### 5.3. Paradigm applications

A prime example of technosignature research per conventional paradigms is the search for signs for artificial constructs on Jupiter's moon Europa. A prime astrobiological target due to its ocean, Europa is scheduled to be studied by the Europa Clipper mission[22], which – while not designed for technosignatures research – will collect two-way Doppler shift measurements using the X-band communication system as part of the radio science package.

Icy ocean worlds like Europa have unique structures that are different from typical terrestrial bodies within our solar system (e.g., Mercury, Earth, the Moon, and Mars). Some consist of a layered-cake internal structure with liquid water bounded by different ice phases (e.g., Titan and Ganymede), whereas others are capped by an ice shell of variable thickness with a subsurface liquid water layer and a silicate interior beneath (e.g., Europa and Enceladus). For Europa, this difference in internal structure corresponds to an exploitable density profile, with regional-scale mass anomalies on the rock-water interface (i.e., impact-volcanic-tectonic structures), and possibly at the base of the ice shell or within the ocean as well (i.e. structures or deposits), producing the largest gravity anomalies observable from orbit after the removal of the contribution from the surface topography[23]. This is due to the stark density difference between ice/water or briny ice and water/rock; however, these data could be attributed to abiotic/artificial materials contained within the internal structure of Europa (e.g., an air-filled, hemispherical dome). Thus, it may be possible to model the strength and spatial distribution of gravity anomalies originating from possible underwater technosignatures and differentiate them from those resulting from geologic processes. Anomalies resulting from near-surface density contrasts could be corroborated with ice penetrating radar (i.e., the REASON instrument) to better constrain their provenance.

Currently, this approach is largely aligned with the parameterized, constrained hypothesis-driven approach, with constraints being placed on the expected gravity data (specifically, on its magnitude and distribution) relative to known artificial substances observed on Earth. However, under the paradigm proposed herein, data is evaluated in a hypothesis-agnostic fashion, with any anomalous phenomena being flagged as potential NCPs and prioritized for further investigation. Subsequently, based on the characteristics of an observed anomaly, investigators have the option of performing a more constrained search for similar phenomena may be performed on existing gravity data, or constructing a dedicated instrument for ascertaining whether the particular anomaly that was observed was attributable to artificial materials or to environmental processes.

In a similar vein, citizen science projects are also engaging in anomaly-based searches for identifying technosignature-relevant NCPs. For instance, the VASCO citizen science project[24] amasses data from astronomical studies and uses a combination of manual anomaly identification by participants and algorithmic anomaly detection methods to flag data that is potentially indicative of NCPs. The emphasis placed by such projects on parsing large quantities of data through several means (with multiple participants reviewing each dataset, in addition to an algorithm) enables a more hypothesis-agnostic search methodology that is well-suited to NCP identification. These types of studies can be paired with the more in-depth characteristic studies (e.g. the Europa Clipper example detailed above) to yield insight into technosignature-relevant NCPs.

### 5.4. Evaluation of lifecycle approach

The proposed investigational lifecycle (where anomaly searches are used to drive more constrained parameterized investigations) is greatly beneficial to NCP-dependent fields such as technosignatures as it greatly minimizes the number of requisite assumptions needed to commence searching for NCPs, and subsequently focuses investigational efforts on potential NCPs that are supported by empirical data (i.e. the identified anomaly or anomalies). As evinced by its compatibility with citizen-science research efforts, the proposed investigational paradigm also better enables NCP-dependent fields to utilize existing observations and analyses from investigations and observational programs from other peripherally related disciplines (e.g. astrophysics). For example, any planetary mission which is imaging surfaces of other solar system bodies at higher resolution than previously obtained could be used to search for non-canonical physical structures. Given that NCPs, particularly in the case of technosignatures, are often expected to be rare phenomena, having a large dataset to parse through is highly advantageous.

Another important consideration in NCP research is the existence of statistical biases (for instance, Malmquist bias in observational astronomy[25], concerning the preferential detection of intrinsically bright celestial objects) which may skew highly constrained analyses. It has already been established in the literature that an ideal methodology to avoid such biases is to analyze a minimally constrained set[26]; while capture and analysis of a wholly unconstrained set is not necessarily feasible, the investigational paradigm presented herein is an optimized approach to minimizing the number of constraints placed during the initial phases of investigation. The advent of algorithms and crowdsourced data parsing programs that are able to handle the large quantities of data relevant to such efforts has also made the implementation of such a paradigm very much feasible.

Thus, whilst narrow-scope, parameterized investigations – the current favored investigational methodology for NCPs – certainly have a place in the study of anomalous phenomena, it is recommended that such investigations in NCP research (such as for technosignatures) be first preceded by broad-spectrum anomaly searches. This dual-phase investigational cycle, now made feasible by the present availability of various rapid data-parsing methods, holds significant promise in the systematic identification of NCPs compared to conventional methods.

**Conclusions**

NCPs are a critical aspect of physical sciences research, and – as is evinced by prior instances – are capable of yielding scientific insights that have value both in validity testing of existing theory and in providing investigational targets for new research. There has been a distinct lack of clear investigational paradigms tailored towards the identification and subsequent characterization of NCPs, with the current favored paradigm – that of heavily parameterized, constrained hypothesis-based investigations – better suited to performing characterization studies than systematic initial identification of NCPs. However, the investigational lifecycle paradigm proposed herein, which utilizes anomaly-driven searches to identify targets for more constrained investigations and secondary studies – is a much more reliable means of capturing data on NCPs in a systematic fashion due to its initial hypothesis-agnostic approach. NCP-dependent fields, such as technosignatures, stand to greatly benefit from the application of this methodology; furthermore, due to the inherent disciplinary nature of NCP searches, such research stands to be of interest to the greater physical sciences community as a whole.

**Acknowledgements**

This study resulted from the TechnoClimes workshop (August 3-7, 2020, technoclimes.org), which was supported by the NASA Exobiology program under award 80NSSC20K1109. Any opinions, findings, and conclusions or recommendations expressed in this material are those of the authors and do not necessarily reflect the views of their employers, NASA, or any other sponsoring organization.

The authors wish to acknowledge the participants in the 2020 TechnoClimes conference for their support and for their role in stimulating the discussion that lead to this paper.

**Author contributions**
CS served as the chair of the discussion session at the TechnoClimes conference which led to this paper, performed the mathematical analyses, developed the figures, wrote the sections not attributed below, and coordinated the paper's development.

JHM contributed the 'Dogmatic context' section.

AB contributed to the 'Case study context' section.

AS (Europa Clipper discussion) and BV (citizen science discussion) contributed to the 'Paradigm applications' section.

DG and GDT contributed the 'Evaluation of approach' section (DG on the introductory portion of the section, and GDT on bias).

HF contributed to the content of the 'Challenges of technosignature research' section.

VT contributed to the content of 'Anomaly-based studies' section.

**Competing interests**
The authors declare no competing interests.